%% file: reciprocity.tex
\documentclass[conference]{IEEEtran}

\usepackage{helvet, courier, graphicx}
\usepackage{subfig}
\usepackage[colorlinks,urlcolor=blue,citecolor=blue,linkcolor=blue]{hyperref}
\usepackage{color}

\ifCLASSINFOpdf
\else
\fi
\begin{document}
%
\title{Dynamics of Trust Reciprocation in Heterogenous MMOG Networks}


\author{\IEEEauthorblockN{Ayush Singhal, Karthik Subbian, Jaideep Srivastava}
\IEEEauthorblockA{University of Minnesota, Minneapolis, MN.\\
\{ayush,karthik,srivasta\}@cs.umn.edu
}
\and
\IEEEauthorblockN{Tamara G. Kolda, Ali Pinar}
\IEEEauthorblockA{Sandia National Laboratories, Livermore, CA.\\
\{tgkolda,apinar\}@sandia.gov\\}
}


%


\maketitle

\begin{abstract}

Understanding the dynamics of reciprocation is of great interest in sociology and computational social science. The recent growth of Massively Multi-player Online Games (MMOGs) has provided unprecedented access to large-scale data which enables us to study such complex human behavior in a more systematic manner. In this  paper, we consider three different networks in the EverQuest2 game: chat, trade, and trust. The chat network has the highest level of reciprocation (33\%) because there are essentially no barriers to it. The trade network has a lower rate of reciprocation (27\%) because it has the obvious barrier of requiring more goods or money for exchange; morever, there is no clear benefit to returning a trade link except in terms of social connections. The trust network has the lowest reciprocation (14\%) because this equates to sharing certain within-game assets such as weapons, and so there is a high barrier for such connections because they require faith in the players that are granted such high access. In general, we observe that reciprocation rate is inversely related to the barrier level in these networks.  We also note that reciprocation has connections across the heterogeneous networks. Our experiments indicate that players make use of the medium-barrier reciprocations to strengthen a relationship. We hypothesize that lower-barrier interactions are an important component to predicting higher-barrier ones. We verify our hypothesis using predictive models for trust reciprocations using features from trade interactions. Using the number of trades (both before and after the initial trust link) boosts our ability to predict if the trust will be reciprocated up to 11\% with respect to the AUC. More generally, we see strong correlations across the different networks and emphasize that network dynamics, such as reciprocation, cannot be studied in isolation on just a single type of connection.

\end{abstract}

\section{Introduction}
The rapid growth in the amount and richness of online interactions, through Massive Multi-player Online Games (MMOGs) such as EverQuest\footnote{\url{https://www.everquest2.com/}} and World of Warcraft\footnote{\url{http://us.battle.net/wow/en/}} are creating social interaction data at an unprecedented scale. These MMOGs help computer scientists and sociologists overcome the key difficulty in studying the social dynamics and human behavior by providing an experimental platform for collecting data at high resolution and for long periods. These virtual worlds provide a rich environment for studying user interactions and have been used in several recent experimental studies \cite{lazer09,Yee06,castronova05,szell2010measuring}. 

\begin{figure}[h]
	\centering
		\includegraphics[width=0.35\textwidth]{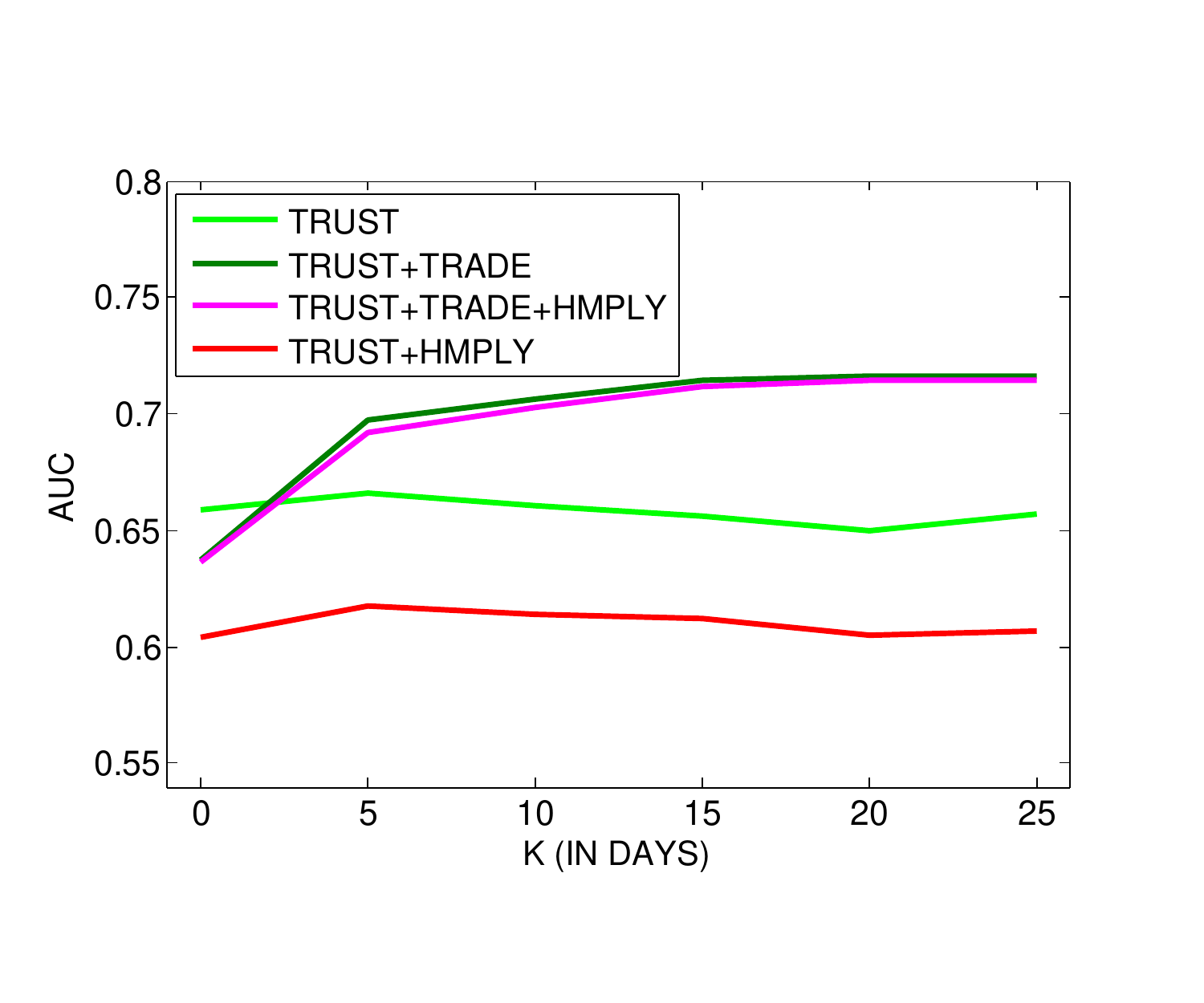}
	\caption{Comparing reciprocation prediction accuracy via AUC. The figure shows that adding hetergeneous network features (e.g. trade) boosts the prediciton accuracy by 11\%. }
	\label{fig:AUC}
	\vspace*{-2 em}
\end{figure}

The ages of the players of these games vary from 13 to 60 and more than 50\% of the players are employed full-time\footnote{\url{http://www.nickyee.com/daedalus/gateway_demographics.html}}. They spend an average of 22 hours per week and 60\% of them reported playing 10 hours continuously\footnotemark[3]. In these games, all user interactions, actions, and communications are recorded in a log file. In this paper, we use one such MMOG called Sony EverQuest II\footnotemark[1], for analyzing the reciprocation of trust relationship in heterogenous interaction networks, including chat, trade, and trust relationships.

The dynamics of the reciprocation varies from network to network depending on the level of barrier for reciprocation. The barrier for reciprocating a trust relationship could be lack of resources or high risk involved. Needless to say, these barriers affect the levels of reciprocation significantly in different networks. For instance, in a chat network users have very low barrier level for trusting each other as there is no commitment from either side to participate in any involved relationship or potential loss. On the other hand, in a trust network players grant access to each other's housing resources. The barrier level in the latter is very high. It is important to understand questions related to reciprocation across different types of interactions. For instance, do people reciprocate differently for trust building activities compared to trust cancellations? As mentioned earlier, understanding these questions in great detail is possible because of the MMOG data. 

The dynamics of complex network relationships cannot be studied in isolation because low barrier interactions may play a critical role in building the high barrier reciprocations. Understanding such dynamics offers several key insights. For instance, our experiments verify that players use low barrier interactions, like trade, before reciprocating trust. We verify our hypothesis by building a predictive model for trust reciprocation; using features from the trade network boosts the AUC by up to 11\% as shown in Figure \ref{fig:AUC}. 

\subsection{Contributions}
The goal of this paper is to understand the dynamics of reciprocation in the context of trust. Following are the key contributions of this work:
\begin{itemize}
\item We explore the relationship between reciprocation and different barriers for reciprocation in several interaction networks. We establish that reciprocation rate is inversely proportional to the barrier level.
\item We analyze the temporal aspect of reciprocation and show that the speed of reciprocation is also inversely related to the barrier level.
\item Our analysis on trust building and cancellation shows that the trust building reciprocations occur more often and more quickly.
\item Reciprocation cannot be studied in isolation, and hence we study the effects of reciprocation in the presence of other interactions. We hypothesize that players use low barrier interactions to establish initial trust, followed by a high barrier reciprocation.
\item We verify our hypothesis using predictive models to find reciprocation links with behavioral features from low barrier networks. Several highly predictive features in general link prediction, such as age homophily~\cite{ahmad10b}, are not as effective as  proposed  heterogenous network features  that boost the AUC by up to 11\%.
\end{itemize}

\begin{figure*}[t!]
\centering
\begin{tabular}{ccc}
\includegraphics[width=.25\textwidth]{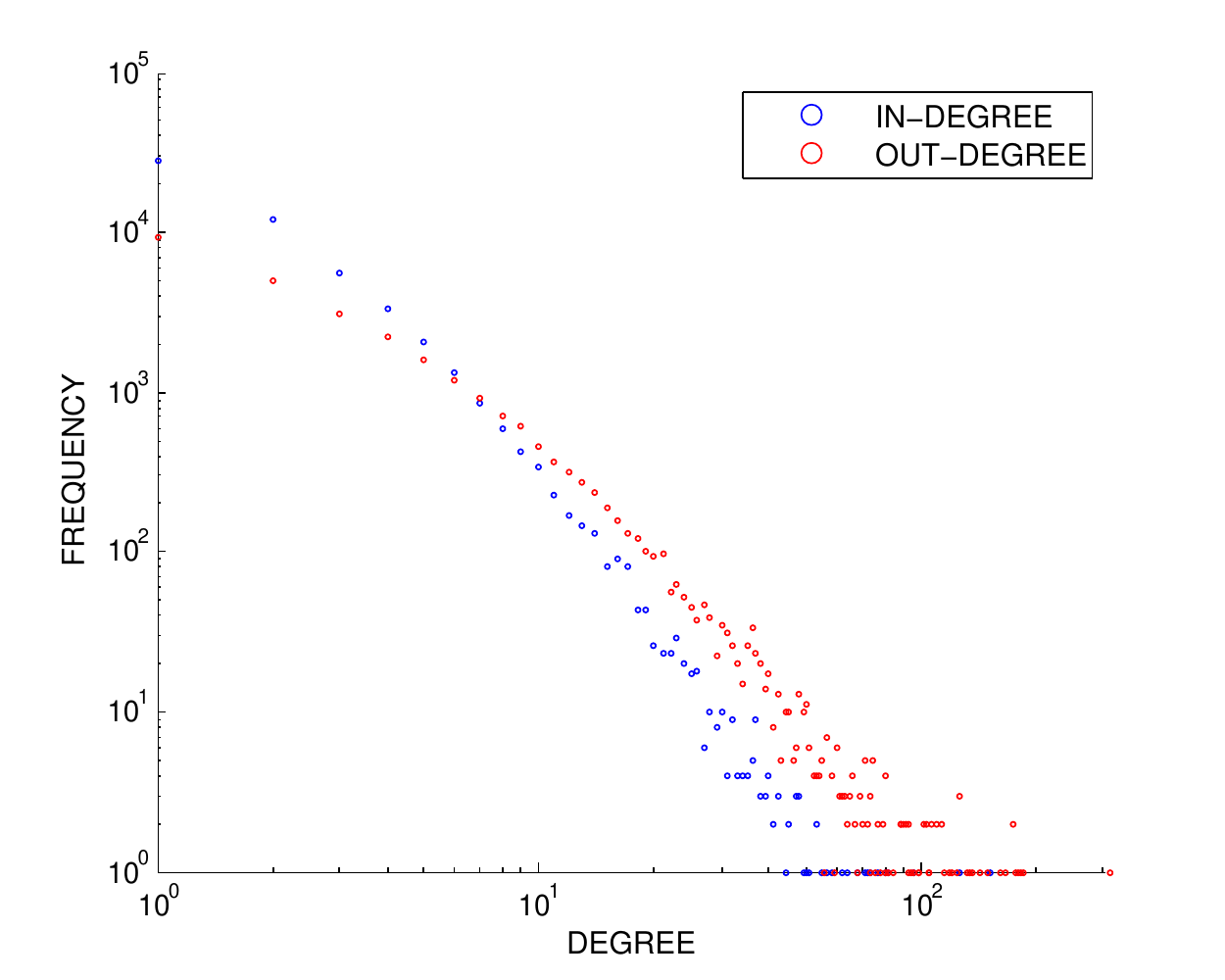}  & 
\includegraphics[width=.25\textwidth]{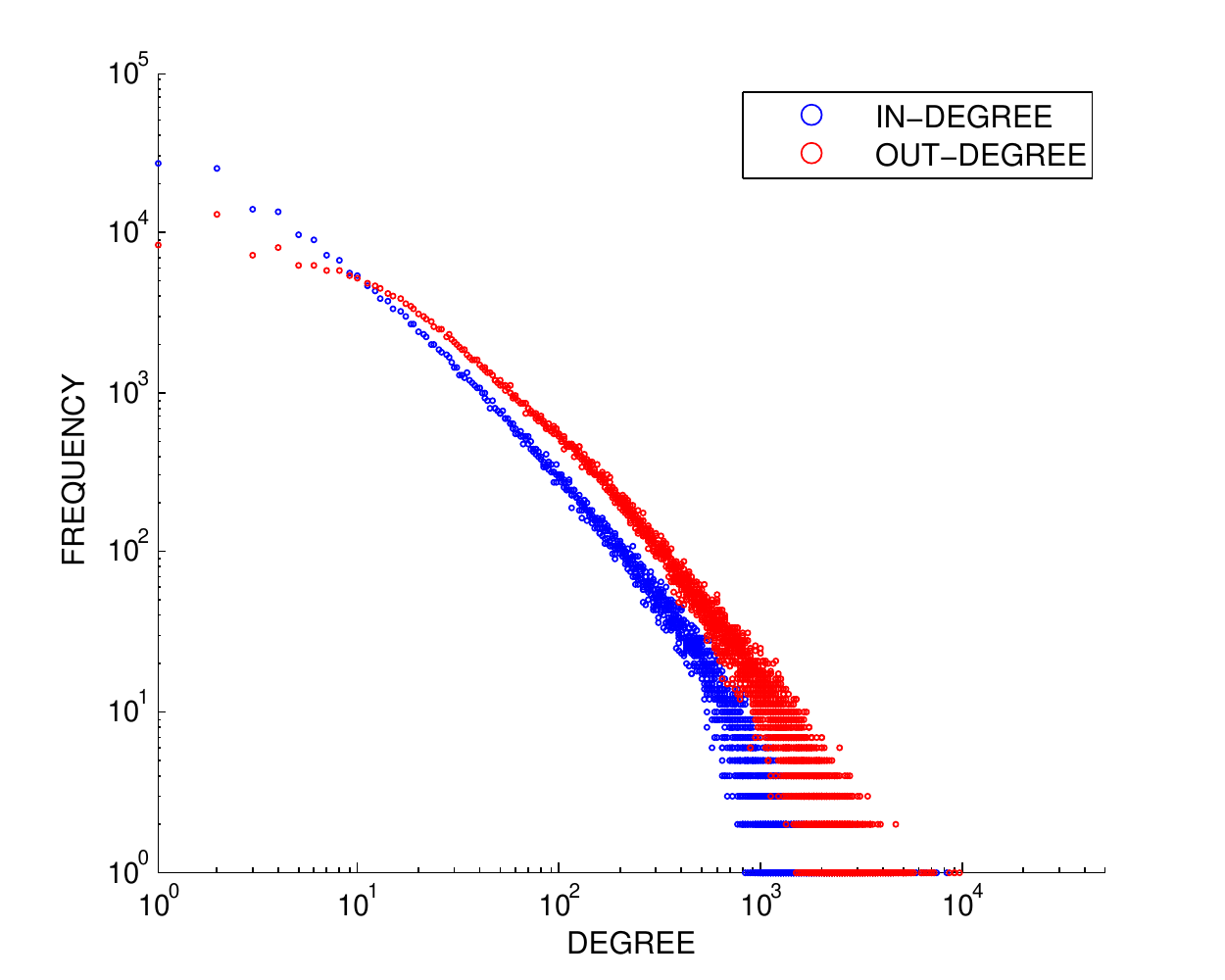}  & 
\includegraphics[width=.25\textwidth]{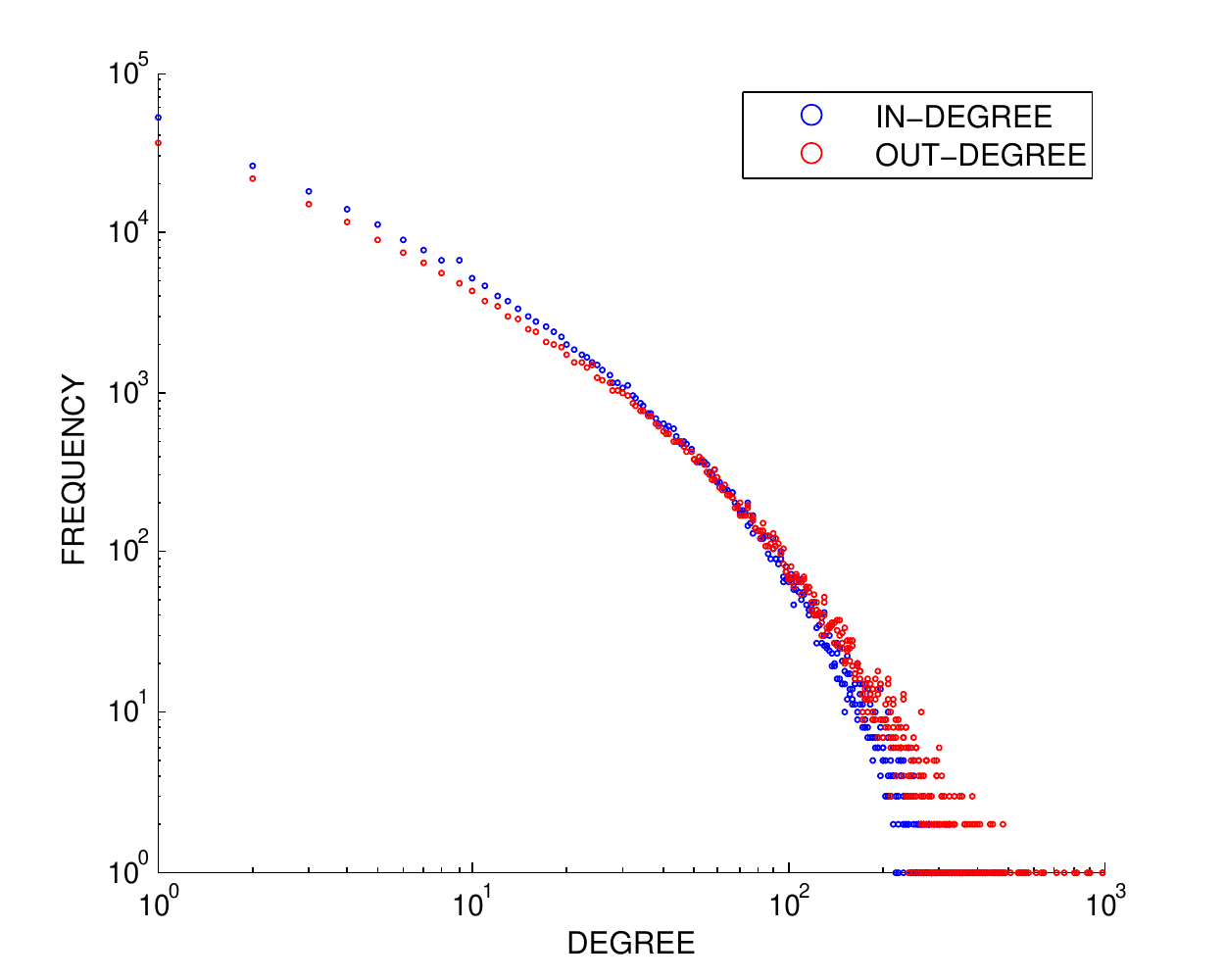} \\
{\scriptsize (a) Trust Network}  & 
{\scriptsize (b) Trade Network} & 
{\scriptsize (c) Chat Network} \\
\end{tabular}
	\caption{Degree distributions of trust, trade and chat networks.}
\label{fig:dd}
\vspace*{-2em}
\end{figure*}

\subsection{Related Work}
The notion of reciprocation is well studied in sociology, and, as defined by Gouldner \cite{gouldner}, it is the norm of reciprocation that people should help those who help them. Indirectly, if you have helped someone then they have an obligation to help you back. Following this notion, several  studies analyze the relationship between reciprocation and the human behavior in various social networks, such as child care \cite{hansen04}, social altruism \cite{leider09}, and online communication \cite{bliss12}. 
Leider et al.~\cite{leider09} use reciprocation as a means to study  human behavior in social altruism in terms of favors and gifting. They conclude, by several online field experiments, that there is a 52\% increase in directed altruism towards friends compared to random strangers. This number further increases by another  24\% when it leads to further prospects of receiving return favors or gifts. Similar experiments were conducted by interviewing families from different working groups to understand the child care favors in social networks. Hansen \cite{hansen04} reveals that social location and kinship improves reciprocation of interaction in these networks. In another recent work \cite{bliss12}, the Twitter reply network is analyzed to understand reciprocation in relation to user happiness, where the measure of happiness is computed using sentiment analysis of text content exchanged in Twitter messages. This work concludes that reciprocal behavior in the Twitter reply network shows high assortativity in terms of measured happiness of Twitter users. In all these works, reciprocation is studied on hand picked user samples or surveys \cite{hansen04,leider09} or used to understand specific human behavioral aspect such as happiness \cite{bliss12} or altruism \cite{leider09}.

There has been recent developments on developing graph models using reciprocation as a model parameter. For instance, Zlatic and Stefancic~\cite{zlatic09} develops a growing model for Wikipedia using three key parameters:  reciprocal edges, degree, and size of modeled network. They empirically evaluate that the generated network closely matches the in- and out-degree distribution of the actual Wikipedia. There are also several papers that discuss the relationship between reciprocation and other network parameters such as degree distribution. Durak et al.~\cite{durak12} propose a null model for directed graphs that combines the reciprocal and the one-way edges to generate directed graphs that match the in-, out-, and reciprocal-degree distributions. Ahmad et al.~\cite{ahmad10a} use theories of social exchange as a basis for building a generative model GTPA for modeling temporally evolving directed networks.

There are other recent works, such as Gralaschelli and Loffredo \cite{gl04} that use statistical measures to conclude that the reciprocation of a network is never at random and it is either reciprocal or anti-reciprocal. They define a correlation coefficient in terms of the average and the actual reciprocation. When the correlation coefficient is  positive and nonzero, it is correlated, and when negative, it is anti-correlated. Similarly, Zamora-Lopez et al.~\cite{zz08} propose a method to compute the expected reciprocation of the network as a  function of in- and out-degree distributions and show empirically the expected reciprocation closely matches the actual in several real-life networks. reciprocation in weighted directed networks, especially in large scale mobile communication networks, is discussed in \cite{akoglu12,wang11}, . Akoglu et al.~\cite{akoglu12}  propose a novel triple power law (3PL) distribution  that fits the reciprocation behavior in several data sets and show interesting properties of this distribution such as parsimony. The paper also concludes that  reciprocation is higher among users with more common neighbors and larger degree correlation. Szell and Thurner \cite{szell2010measuring} study the strength of positive and negative ties in friend and enemy networks and propose approximate social laws relating betweenness centrality and communication strength to the number of overlapping links. In another work Szell et al. \cite{szell2010multirelational} show that negative interactions have a lower reciprocation compared to the positive interactions. 

Reciprocation has also been used for link prediction~\cite{cheng11,zlatic2009,ahmad10b}. Cheng et al.~\cite{cheng11} study several attributes for predicting reciprocation in the Twitter network, such as relative degree, absolute degree, and other link prediction features. They use regression and decision tree classifiers to predict the expected reciprocation and conclude that features that measure the relative status of two nodes  are the best predictors of reciprocation. Zlatic and Stefancic~\cite{zlatic2009} study how reciprocal edges influence other properties of complex networks, such as degree distribution and correlations. They present a statistical inference technique to estimate the number of reciprocal edges. 

In summary, all these related works address the following class of problems: modeling a generative process incorporating reciprocal edges, understanding the effect of reciprocation on a particular human behavior, or using reciprocation-based features in the task of link prediction. None of these papers attempt to study multiple relational networks such as trade, chat, etc. to understand the notion of reciprocation in trust relationship in heterogeneous networks, especially on large-scale real-life networks such as MMOGs. Also, these models do not use an evolving network scenario, where understanding the behavior of reciprocation on the time dimension is extremely important.

\section{MMOG Data Sets}
The Sony EverQuest (EQ) II game provides an online environment where multiple players can log in and coordinate with each other to achieve a particular mission. Note that players are free to invent, choose their mission and to self-organize among groups of their own interest. The game provides several mechanisms such as chat for instantaneously interaction. We were provided with the game data set logs, and we extracted the information needed for our experiments from these logs for various interactions. In this section, we summarize each of these networks in terms of the number of nodes and edges, the period of observation, and the direction of edges. Also, some networks such as trust are rich in terms of edge attributes, we discuss them in more detail.

\subsection{Trust Network}
In EQ II, players form teams in order to complete the game tasks. As the players are limited by the number of items they can carry at a time, players buy houses as a temporary storage to retain their armory and other accessories. Through the trust network, players share their house access with other players. For this reason, we also refer to this network as the trust network. We have 9 months of data from Jan-01-2006 to Sep-11-2006 with 63684 nodes and 140514 edges. Each node in the network is a player character in the game, and each edge is a permission granted or removed by the character to another character. Each edge has a time stamp when the access was granted or removed. In addition, each edge also has a trust level: {\em Trustee, Visitor, Friend, None,} and {\em Remove}. The {\em Trustee} access is the highest level of trust, whereas {\em Remove} is the lowest level of trust the player can express towards another player. The number of edges of the network by trust level is summarized in Table \ref{tab:trustEdges} and each trust level is described as follows.

\begin{itemize}
\addtolength{\itemsep}{-0.5\baselineskip}
	\item {\it Trustee}: Player can store, touch, move, add, and remove things, and has almost same access as the owner.
	\item {\it Friend}: Player can store, touch, and move things.
	\item {\it Visitor}: Player can enter the house and view things.
	\item {\it None}: Player can see the house externally but cannot enter it. 
	\item {\it Remove}: Permission granted to the player is revoked and the house is visible only when the player is near-by.
\end{itemize}

\begin{table}%
\caption{Number of edges by each trust level.}
\label{tab:trustEdges}
\begin{center}
\scalebox{0.9}{
\begin{tabular}{|c|l|c|}
\hline
{\bf Trust Level} & {\bf Description} & {\bf Count}\\
\hline
4&Trustee&82498\\
3&Friend&30850\\
2&Visitor&10928\\
1&None&4336\\
0&Remove&11902\\
\hline
\end{tabular}}
\end{center}
\vspace{-2em} 
\end{table}

\subsection{Trade Network}
In the EQ II trade network, players exchange goods for coins or  goods. Players trade multiple items and coins with other players by  initiating a trade offer. A typical offer includes items/coins offered,  and items/coins needed in exchange. The player offered can either accept or reject the trade request instantaneously. If the request is accepted, then a trade link is established between the seller (initiator) and the buyer (acceptor) in the trade network. We analyzed such a trade network containing 295,055 nodes and 11,913,994 edges over a period of 9 months from Jan-01-2006 and Sep-11-2006. The trust needed to establish a trade relationship is much lower than housing, as the period of interaction is short and the risk involved is nill.
\begin{table*}[t!]
\caption{Statistics of reciprocation in trade, chat and trust networks. The forward edge count also includes the count of multiple forward interaction between all pair of nodes.}
\label{tab:reccount}
\center
\begin{tabular}{|l|c|c|c|c|c|c|}
\hline
Network & All Forward & First 	 & Second 	& Third			& All Other & Total \\
Type (period) 	& Edges 	& Reciprocation & Reciprocation & Reciprocation & Reciprocation & Reciprocation \\
\hline
\hline
Chat  (1 month)	& 1840492 & 441039(23.9\%) & 79412(4.3\%) & 32128(1.7\%) & 46969(2.6\%) & 599548(32.6\%)\\
Trade (9 months)	& 520861	& 74137(14.23\%)	& 11850(2.3\%)	& 3766(0.72\%)	& 47056(9.0\%) & 136809(26.3\%)\\
Trust (9 months) & 62674   & 8452 (13.5\%) & 351 (0.56\%)  & 0(0.0\%)           & 0(0.0\%) &  8083 (14.0\%)\\
\hline

\hline
\end{tabular}
\end{table*}

\subsection{Chat Network}
The chat network is a communication medium  where players exchange instant messages. This network is directed and the direction of an edge is from the sender to the receiver. As the number of chat messages between two players is quite high, and there is no associated session details, we assume all the chat messages in a single day correspond to a single session. The number of nodes in this network is 349,654, and the number of edges is 86,948,748, spanning over a period of one month from Jul-29-2006 to Sep-10-2006. 

\subsection {Network Profiles}
We present the degree distribution of these networks in Figure \ref{fig:dd}. The distributions were constructed using snapshots of different networks over the entire observation period and we considered multiple interactions between a pair of players for this plot. The distributions seem to follow the power law with exponent of the power law ranging from 1 to 3. The exponent was calculated using the slope of a least squares fit in the log-log plot. Now, we look in to the much detail of these networks to understand how players reciprocate in various networks in a snapshot and over time. 

\section{Reciprocation in Different Networks}
The reciprocation of a network is the ratio of  forward edges (say from player $a$ to $b$) that have a corresponding backward edge (from player $b$ to $a$), i.e.,  the ratio of mutual interactions. 

\subsection{Barriers of Reciprocation}
Barriers of reciprocation  can be broadly grouped in to risk and utility. The risk factors include loosing an asset, in-game points, or in-game time; and the utility includes, immediate gains in terms of points and assets and long term future prospects. Each network has specific characteristics that lead to low, medium or high barriers for reciprocation.

The {\em trust network} yields direct access to a player's house. This is the {\em highest} level of barrier as the other person can enter, add, and remove assets from a player's house. In the case of trade network, coin and items are exchanged between the players on a pre-specified contract, so the barrier is having sufficient items or coins for trade. The resulting risk involved is nill with immediate utility for both players. Such networks fall in to {\em medium} barrier category. In the {\em chat network} players are at no risk and there is minimal cost to reciprocation, so these networks fall in to the {\em low barrier} group. For the purposes of this paper, we limit ourselves to one example for each group for our analysis.

\subsection{Multiple reciprocations}
\begin{figure}%
\includegraphics[width=0.9\columnwidth]{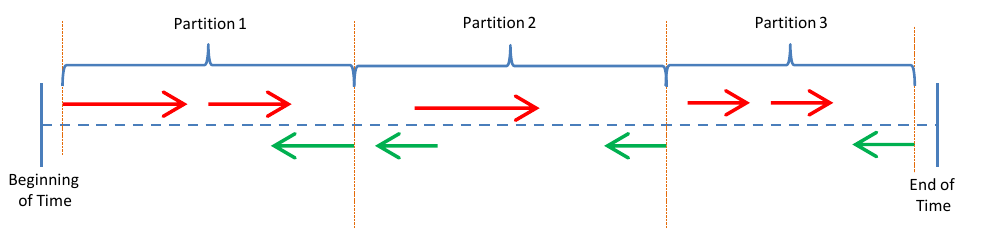}%
\caption{The timeline showing the first, second, and third partitions used for computing the response times for first, second, and third interactions.}%
\label{fig:illus}%
\vspace*{-2em}
\end{figure}

There can be several overlapping forward and backward arcs between each pair of players. For the purpose of measuring the reciprocation and response time, we first partition the timeline into several partitions. We consider the start time of first forward edge and the corresponding end time of the first response as the first partition. Similarly, the second forward arc and its response marks the end of the second partition and so on. This notion of partitions of the timeline is illustrated in Figure \ref{fig:illus}. The figure shows the forward arcs above the timeline arrow (dashed line) and backward arcs (reciprocations) below the timeline arrow. For remainder of this paper, unless specified, we always refer to the overall reciprocation rate. In Table \ref{tab:reccount} we show multiple reciprocation rate for two different networks.

\begin{figure*}[ht]
       \centering 
        \subfloat[Trust] {\label{fig:granting}
         \includegraphics[width=0.3\textwidth]{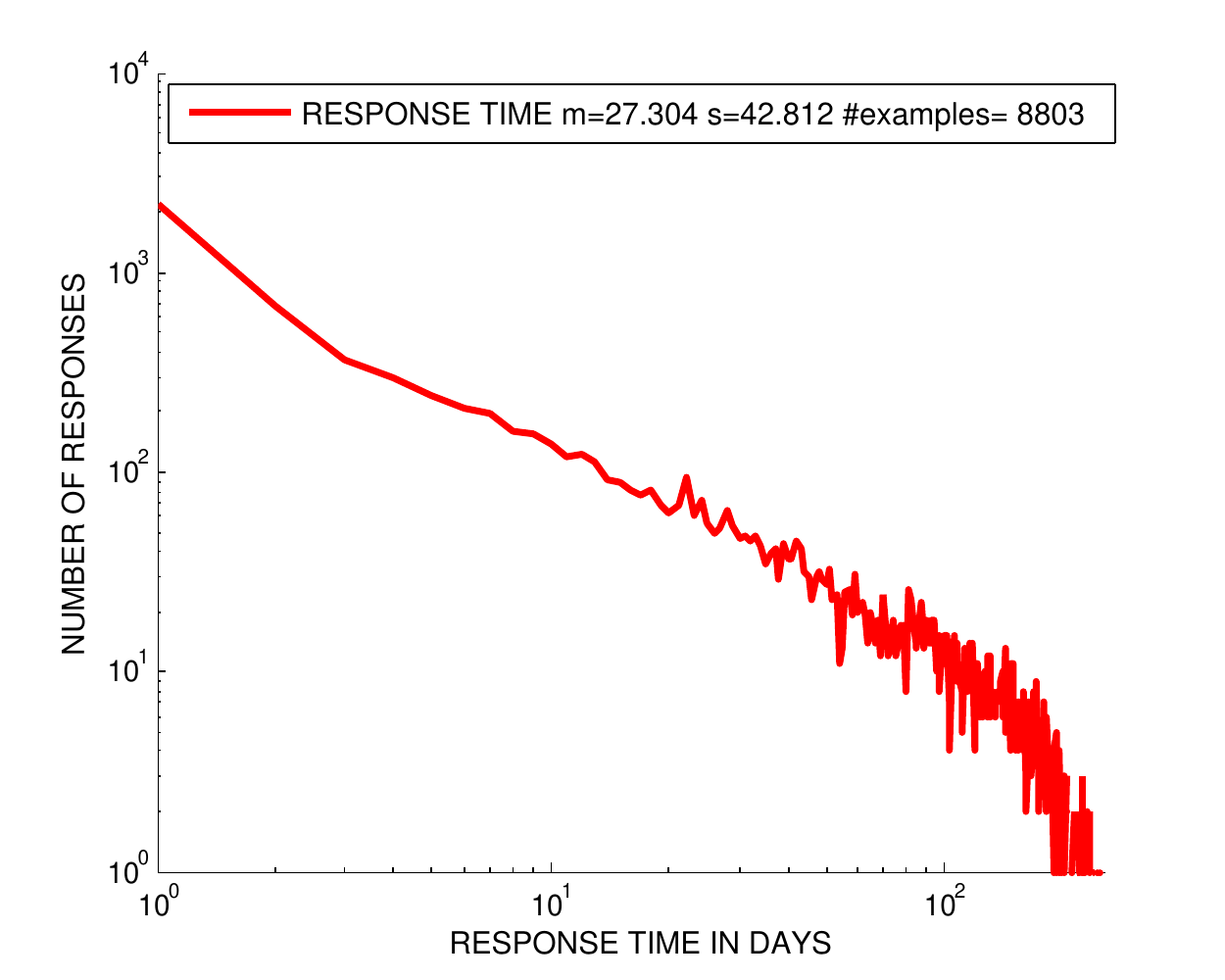}
         }
          \subfloat [Chat] {\label{fig:chatall}      
           \includegraphics[width=0.3\textwidth]{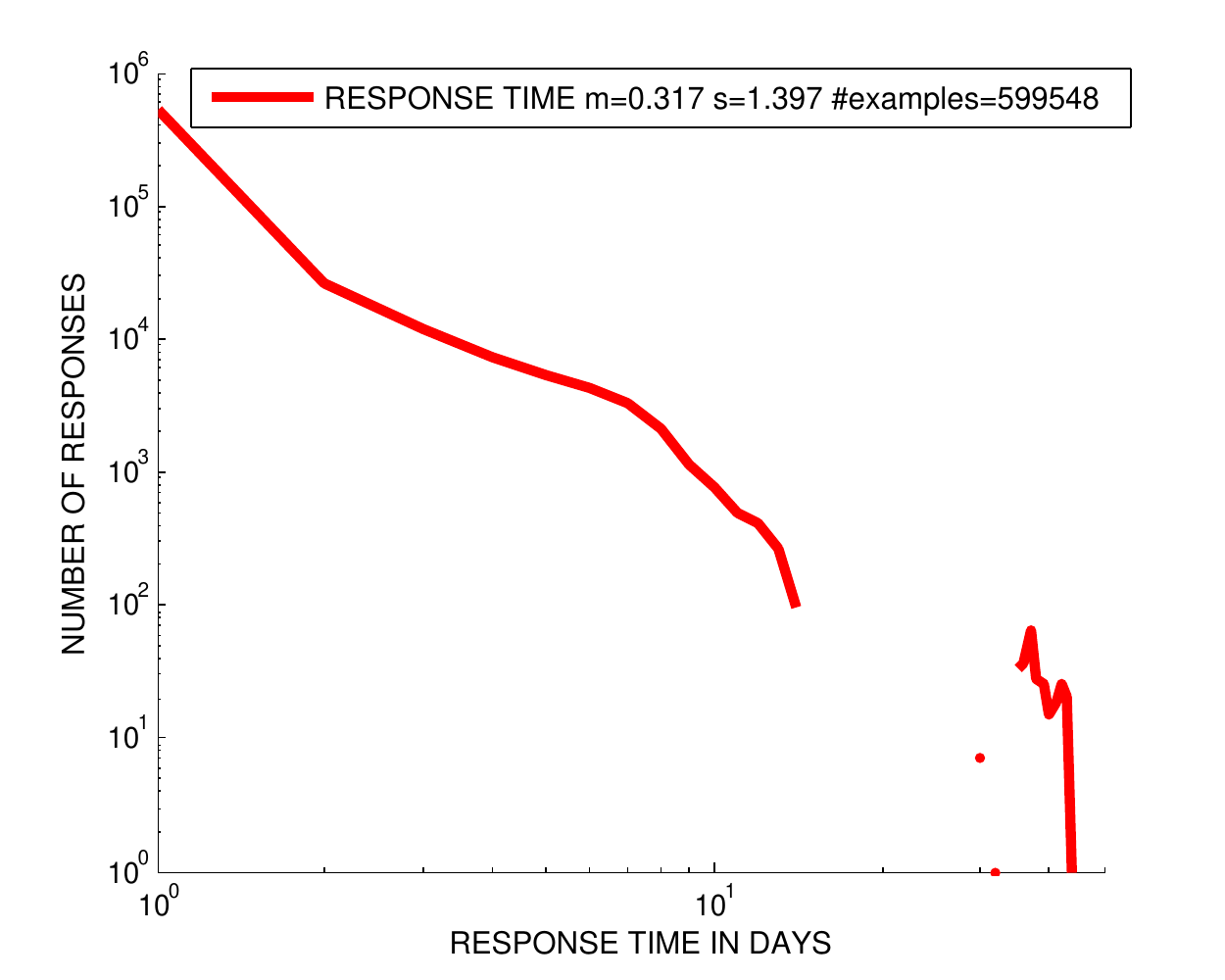}
             }
             \subfloat[Trade] {\label{fig:tradeall}
                \includegraphics[width=0.3\textwidth]{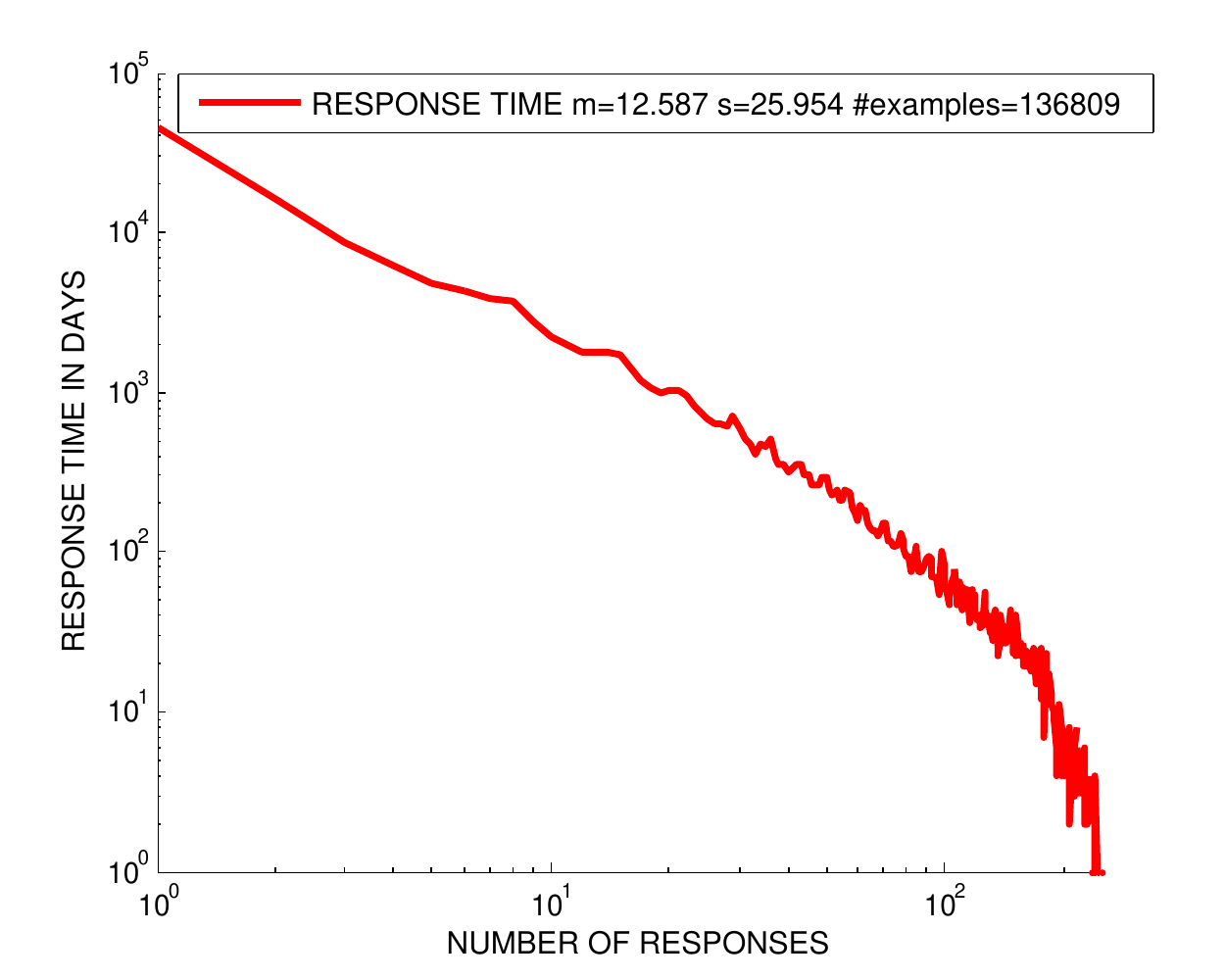}
             }
            
             \caption{The response time distribution  for the three networks. The average response time for trust network is the highest among chat, trade and trust.}
             \vspace*{-2em}
\end{figure*}

\subsection{Reciprocation in Trust Network}
The trust network is a high barrier reciprocation network as players have to trust each other in terms of granting house access. It consists of several levels of access permission: Trustee, Friend, Visitor, None, and Remove. As we see in Table \ref{tab:trustEdges}, the percentage of edges that belong to the Trustee category accounts for more than 50\% and some categories, like None, are rarely used. Also the dominant transition is from other permissions to Trustee and most of these permissions remain in Trustee state until the end of the observation period. For these reasons, we assume that there are only two major permission categories in the trust network: {Trust} and {Not-Trust}. The Trust level corresponds to being a Trustee. We collapse the lower trust levels, ({Freind, Visitor, None, and Remove}), to a single {Not-Trust} level. As the lower level permissions are equivalent to having no relationship at all, we consider absence of an edge as the {Not-Trust} state. In essence, we focus only on the Trustee level of access and all the other links are as equivalent to having no access at all. 

In the trust network, 14.0\% of the forward trust (8803 one-way) links receive a trust response (reciprocation) back and their response time distribution is shown Figure \ref{fig:granting}. We study the response time between the reciprocations and see how quickly and in what volume these reciprocations happen using the response time distribution. We analyze the entire graph edge wise for this purpose. If an edge occurs from node $i$ to $j$ at time $t_1$, then we will account for the start time of that edge as $t_1$. If the reciprocation occurs from node $j$ to $i$ at time $t_2$ then the response time is computed as $t_2-t_1$. All the forward edges from $i$ to $j$ between $t_1$ to $t_2$ are subsumed by the first forward edge, until a reciprocation response is received by $i$ from $j$. We then use this data to draw the distribution of response times. The x-axis denotes the time to reciprocate and y-axis corresponds to the number of reciprocations. We find that the response time distribution follows a power law, with a mean response time of 27 days. 

We also measured the number of player who waited indefinitely for a response back from the other player until the end of observation period. Surprisingly, 68.54\% of players waited indefinitely, while the remaining 8.16\% of players switched from Trust to Not-Trust state as they did not receive a Trust response. This indicates that around 69\% of players in the trust network show patience.

The users, in the trust network, can move from Trust to Not-Trust state and at a later date they can again establish a new Trust relationship. As this network has an unique feature of canceling the edges, we wish to study the reciprocation of granting trust, followed by cancellation. We measure the number players who initiate a revocation of housing permission and the number players who reciprocated by a similar action. From the 8803 reciprocated links 1062 (12.1\%) of them initiated a reduction of housing access to their peer player and 207 players out of 1062 (19.4\%) reciprocated back with a revocation response. About 80\% of the players who received a cancel request never responded back and ignored the  request. This again confirms that when a reciprocation relationship has high barriers, an established mutual trust is quite stable to cancellation from one-side. Also, the players seem to respond to trust requests with a mean response time of 27 days compared to a cancellation request of 32 days. This suggests that the players pay more attention to granting requests compared to canceling requests, as establishing trust is a more preferred social activity than losing it.

\subsection{Reciprocation in Chat,Trade and Trust Networks}
In this section, we study the reciprocation behavior in chat and trade as independent homogenous networks. Then, in the next section, we analyze the interactions of chat and trade relationships in affecting the reciprocation of trust network. The chat and trade networks have simple edge types with no attributes except time stamps. 

We summarize the total reciprocation for each network type in Table \ref{tab:reccount}. As the chat network is a low barrier network, it has the highest amount of reciprocation with 32.6\% of forward edges reciprocated. The low barrier in this network is due to the instant nature of the communication and minimal risk involved in making a chat reciprocation. On contrary, in the medium barrier trade network the reciprocation rate is 26.3\%, lower than chat but more than the trust network. There several reasons why a player may not reciprocate in this network, it could be either lack of resources or a need for doing so. Further, we wish to analyze the response time distributions of these networks to understand some key questions, such as, does all reciprocations occur within a certain number of days or are they spread uniformly over a longer period of time?

Figure \ref{fig:chatall} shows  the response time distribution, which  roughly follows a power law. There is an outlier region around 45 days. We investigated this region and found that these are first time users who are not familiar with the system. We note that such users are extremely rare in the dataset (less than 0.01\%). The figure also shows that most of the users in the low barrier, chat network reciprocate within the same day or at most the next day. This is evident since the mean first response time in the chat network is less than one day (0.317). In the chat network there is a sharp truncation \cite{amaral00} after 7 days, as the significance of a message beyond a week becomes completely irrelevant to the context of the game.

For the trade network, we show the response time distribution in Figure \ref{fig:tradeall}; the distribution has a heavy tail and seems to  follow a power law. The slope of this distribution is not as steep as for the chat network, implying the trade reciprocation is not as quick as  in chat. As the barrier for reciprocation in the trade network is more than chat, the average response time in trade is 43X slower at around 13 days.

\input{heterogenous}
\section{Conclusion}
Understanding reciprocation in networks is important in several fields from computational social science to sociology. In this paper, we have extensively studied various social factors affecting reciprocation in three different interaction networks from Sony EverQuest II MMOG. These networks vary in trust levels, interaction type and sparsity and give a broad perspective of how players reciprocate across a variety of networks. We established the connection between the trust level required to establish a connection inversely affects the number of reciprocations. We also, using response time analysis, show that people are slow in building mutual high trust relationships compared to low trust ones. We also show that in high trust networks player patience plays a key role in reciprocation. We extend our analysis from single-type networks to heterogeneous networks, where we confirm that high degree of socialization is crucial for reciprocation in high trust relationships.

\section*{Acknowledgments}
This work was funded by the GRAPHS Program at DARPA. Sandia National Laboratories is a multi-program laboratory managed and operated by Sandia Corporation, a wholly owned subsidiary of Lockheed Martin Corporation, for the U.S. Department of Energy's National Nuclear Security Administration under contract DE-AC04-94AL85000.

\bibliographystyle{IEEEtran}
\bibliography{ref}

\end{document}

%% file: heterogenous.tex
\section{Reciprocation in Heterogeneous networks}
We will now overlay the trust, chat, and trade networks to analyze the interactions among heterogeneous edges. As the chat data is available only for a month, we restrict the other data sets also to this one month period. Every pair of players in this network have several edges between them with time stamps for each edge denoting when the edge occurred. For the trust network we consider the TRUST and NOT-TRUST edges. The focus of this section is to answer questions, such as: How many times does a trust granting from player $a$ to $b$ result in a trust reciprocation from $b$ to $a$? Does a player $b$ prefer to reciprocate with trade or some other low barrier interaction such as chat before granting a high barrier relationship such as trust to player $a$.
 The first interaction, as noted earlier in Figure \ref{fig:chatall}, captures all the characteristics of the additional interactions, hence we consider only the first interaction for this experiment. 
For each forward edge type, we count the number of first reciprocation between pairs of players in the consolidated network. In the case of tie reciprocations across multiple edge types we include all the tied edges while counting.
We have summarized the reciprocations for each forward interaction type in Table \ref{tab:precur}. 

\begin{table}%
\caption{The reciprocation counts for first interaction (first forward request and first reply) in a heterogeneous MMOG network for a period of one month. The forward edge count includes only the first forward edge between any pair of nodes. The reciprocation captures the first type of response (chat, trade or trust) to the forward edge.}
\label{tab:precur}
\center
\begin{tabular}{|l|c|c|c|c|}
\hline
Forward & First Forward & Chat & Trade & Trust \\
Type & Edges 	& Reciprocation & Reciprocation & Reciprocation \\
\hline
\hline
Chat & 1645623 & 435758 & 1187 & 105 \\
Trade & 74428 & 7953 & 11402 & 335 \\
Trust & 10502 & 907 & 1016 & 722 \\
\hline
\end{tabular}
\end{table}

As we see from Table \ref{tab:precur}, the players who perform chat predominantly make a reciprocation using the chat link (26.48\%). The main reason for this is that a chat relationship is a low barrier relationship, which can be established between two players instantly and does not require high degree of trust.
 For this reason, players do not hesitate to reciprocate a chat response as the amount of cost involved is insignificant and the utilities are never worse off. This observation justifies the theory of rational selfishness \cite{baier90}, whereby it is in the best interest of the player to reciprocate to a chat request to maximize his/her utilities. The same observation is made for the trade relationship, which is also a low barrier relationship (but higher than chat). Here also we find that trade responses are the predominant type of responses for a trade relationship. This is in line with bargaining theory \cite{rhodes89} where bilateral parties who often negotiate a successful deal tend offer favors to each other again. However,this is not the same in the case of a trust forward edge type as it is a high barrier relationship and a more time consuming activity. So people are very careful before reciprocating for such activities and reciprocations are first initialized through low barrier activities before reciprocating with a high barrier relationship, such as trust. The reciprocations for trust forward edge is predominantly through chat (8.63\%) or trade (9.67\%).

We now analyze the dependency between trust interactions against trade and chat interactions. Recall that trade is often a high barrier relationship whereas chat and trade are instantaneous and hence have a low barrier. The aim of this experiment is to quantify how chat and trade, low barrier relationships, influence reciprocation in a high barriermrelationship. 

For any two nodes $a$ and $b$ in the trust network, we start our analysis from the time when a forward TRUST edge from $a$ to $b$ is established. For such a forward edge there can either be a TRUST reply to complete the TRUST relationship or no TRUST reply (incomplete TRUST relationship). The TRUST relationship is determined as incomplete if there is no trust response from $b$ to $a$ within the average trust response time which is 4.6 days in this case. In other words, we truncate the response time for the incomplete reciprocations by the mean response time (4.6 days) and use only the period before this mean response time for further analysis. There can also be several other responses (low barrier interactions) from $b$ to $a$ before $b$ replies with a TRUST link. Understanding these other relationships, such as chat and trade, before a TRUST reply is formed from $b$ to $a$ is crucial to decipher the nature of socialization required for a healthy mutual trust relationship.

From Table \ref{tab:new_precur} we can infer that complete and incomplete TRUST rows differ from each other in terms of chat and trade responses. We can see that the responses are exactly in an opposite order in the two rows. For the complete TRUST we observe that there were 743 forwards edges that were responded back with TRUST.  However, before the TRUST is completed between $a$ and $b$, we see that there are 408 trade responses from $b$ to $a$. These trade responses account for nearly 63\% of the total responses. We have only 243 chat responses from $b$ to $a$ , which is comparatively smaller than the trade responses. Surprisingly, the amount of low barrier responses for the incomplete TRUST gets completely reversed. There are 9145 forward TRUST edges which do not get a TRUST reply back and remain one sided. For these TRUST requests from  $a$ to $b$, we find that the chat responses from $b$ to $a$ are now 6962 (approximately 75\% of the total responses) while the trade responses are significantly lower (as low as 25 \%). 

This experiment confirms that a TRUST relationship between $a$ and $b$ is more likely to complete if there are more trade responses than chat responses from $b$. This interesting result can be used to infer the future TRUST relationships based on some low barrier relationships such as chat and trade.

\begin{table}[h]
\caption{The effect of low barrier relationships on high barrier relationship (TRUST). This tables shows the connection between chat and trade reciprocation counts and completion or incompletion of trust reciprocation.}
\label{tab:new_precur}
\center
\scalebox{1.0}{
\begin{tabular}{|l|c|c|c|}
\hline
TRUST & Forward & Chat &Trade \\
Type & Edges & Responses  & Responses\\
\hline
\hline
Complete & 743 & 243 (37\%) & 408 (63\%)\\
Incomplete & 9145 & 6962 (75\%) & 2331 (25\%) \\
\hline
\end{tabular}}

\end{table}

\section{Predicting trust reciprocation}
In this section we evaluate how well can we predict a high barrier relationship, such as a trust, using information about the medium barrier relationships between the nodes. The empirical analysis in the previous section showed that the success (completion) of high barrier trust relationship depends on some medium barrier relationship like trade. We use this as our motivation to quantify how well the medium barrier relationship can help to predict high barrier relationship completion. However, we use the entire 9 months of data in order to make any conclusions for the trust reciprocation prediction. The chat relationship has to be excluded from this experiment because of its limited availability for a single month. But we add several other features to make the experiment more interesting.

For the trust network data for 9 months, there are a total of 61006  trust requests (forwards edges).  The trust links that are reciprocated is 8252 whereas 52574 forwards edges remained incomplete. For the forward requests, we ignored the requests that started in the last (9th) month because it is hard to determine whether the requests were completed. Thus the number of trust requests are slightly lesser than those mentioned in earlier sections. So the completed trust link (reciprocated) forms one class and the incomplete trust links (unreciprocated) forms the second class. Now that we have two classes, we construct several features in order to build a prediction model. The following features will be used to build prediction models.

\textbf{Features from high barrier relationship(trust)}: This set of features are built using the structural properties of the trust network. These features characterize the position of players (nodes) in the trust network. For the reciprocation links $(A,B)$ we consider  two structural features. The first structural feature describes the connectivity of $A$ to other nodes in a trust network. The second structural feature is the connectivity of $B$ with other nodes in the trust network. For convenience, we refer these features as ``trust" features. 

\textbf{Features from medium barrier relationship(trade)}: This feature set consists of features from three sub-categories namely, structural, past-behavioral and future-behavioral. For a link $(A,B)$ the structural feature corresponds to the degree of $A$ and degree of $B$ in the trade network. The past-behavioral features for a link $(A,B)$ correspond to the count of the trade interaction of the type $A$ to $B$ and $B$ to $A$ before the trust request from $A$ to $B$ started. This feature takes into account the trade behavior between $A$ and $B$ before any trust interaction started between them. The future-behavioral features takes into account the behavior of trade interaction between $A$ and $B$ once the trust request is sent from $A$ to $B$. Here we use a time window  $K$ (in days) starting from the time when trust request was initiated from $A$ to $B$. We count the number of trade interactions in this time window $K$.

\textbf{Features from player demography(homophily)}: In this feature set we take into account the two types of homophilies. The first type of homophily is gender homophily. The gender homophily between $A$ and $B$ is 
1 if $A$ and $B$ has the same gender, and 0 otherwise. 

The second demographic feature is the experience homophily between $A$ and $B$. It is computed as  $X(A,B)=X(A)-X(B)$, where, $X(A)$ is the experience level of $A$. 

As mentioned earlier, the aim of this experiment is to quantitatively compare the impact of using different features (described above). In the previous section we hypothesized that the success of trust reciprocation (completion) can be determined by the amount of medium barrier interactions between player $A$ and $B$. Thus to validate this hypothesis, we conduct several experiments using different set of features.

\begin{table}
	\centering
	\caption{Table comparing reciprocation prediction accuracy using different feature sets. Here K denotes the time window starting from the time when an initial trust link was established between two nodes.}
	\scalebox{0.9}{
		\begin{tabular}{|l|c|c|c|c|c|}
\hline
& & & Average & Average & \\
Classifier & CWA & AUC &  Precision &  Recall & F-measure\\

\hline
\hline
only trust & 0.515 & 0.659 & 0.800 & 0.863 & 0.806 \\
trust+trade(K=0) & 0.526 & 0.637 & 0.825 & 0.866 & 0.816 \\
trust+homophily & 0.519 & 0.604 & 0.788 & 0.849 & 0.808 \\
trust+trade(K=0)+homophily & 0.527 & 0.636 & 0.826 & 0.866 & 0.817 \\
trust+trade(K=20) & 0.588 & 0.714 & 0.871 & 0.885 & 0.851 \\
\hline
\end{tabular}}
	
	\label{tab:TableComparingReciprocationAccuracyUsingDifferentFeatureSets}
	 \vspace*{-2em}
\end{table}

\begin{figure*}[t]
       \centering 
        \subfloat [Class weighted accuracy] {\label{fig:cwa}
         \includegraphics[width=0.4\textwidth]{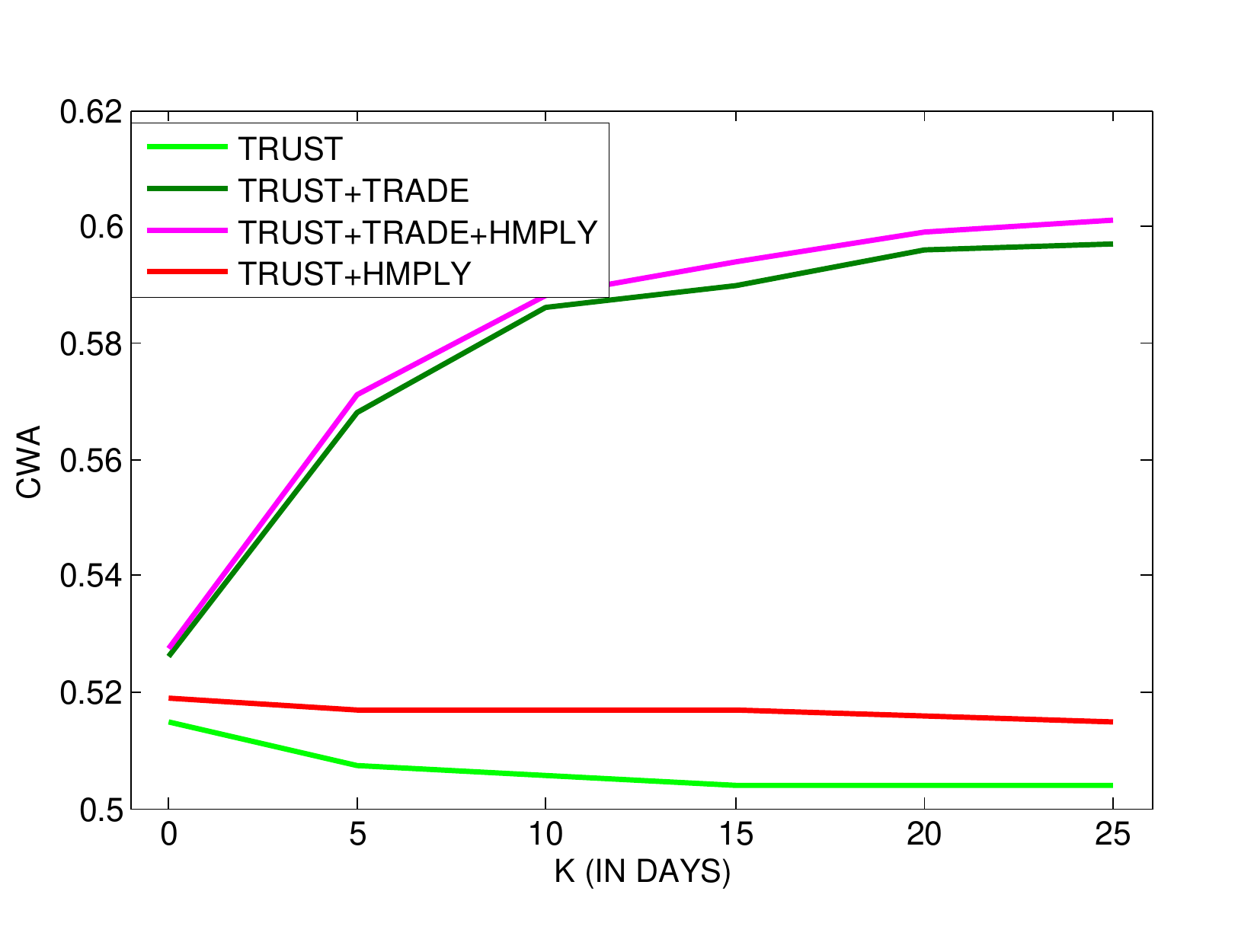}
         }
          \subfloat[$F_{1}$ measure] {\label{fig:fmeasure}      
           \includegraphics[width=0.4\textwidth]{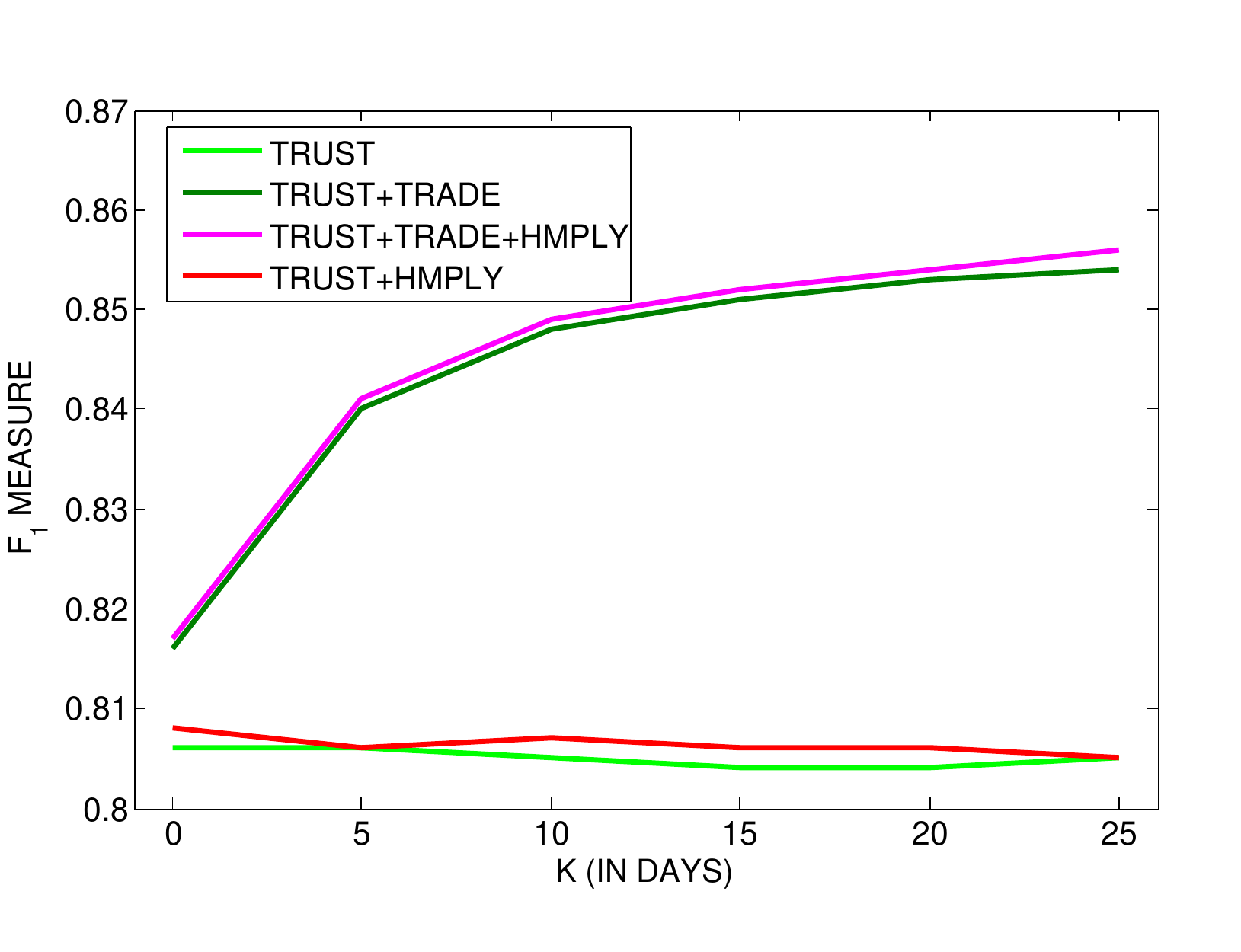}} \\
             
             \subfloat[Precision]{\label{fig:precision}
                \includegraphics[width=0.4\textwidth]{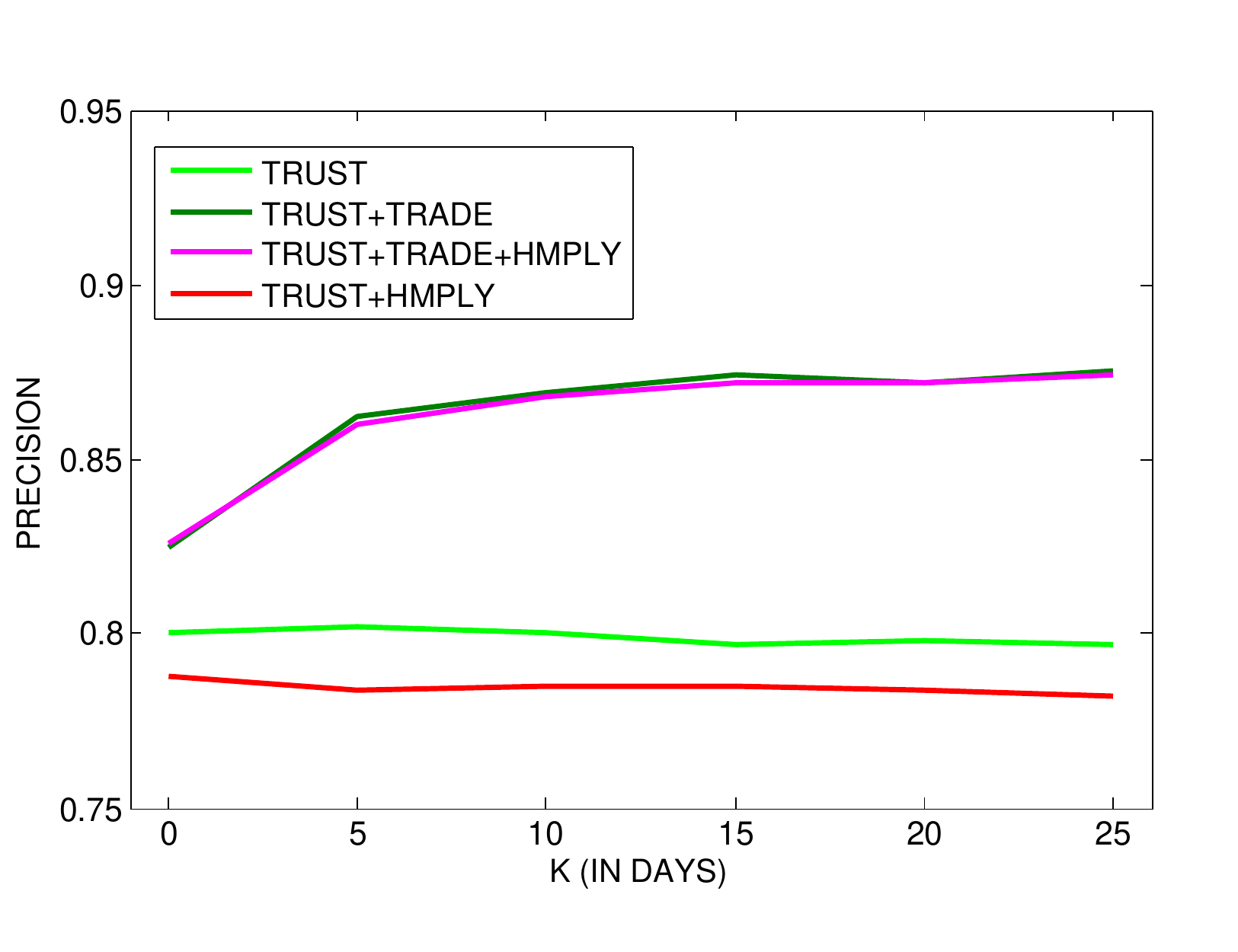}
             }
              \subfloat[Recall]{\label{fig:recall}  
              \includegraphics[width=0.4\textwidth]{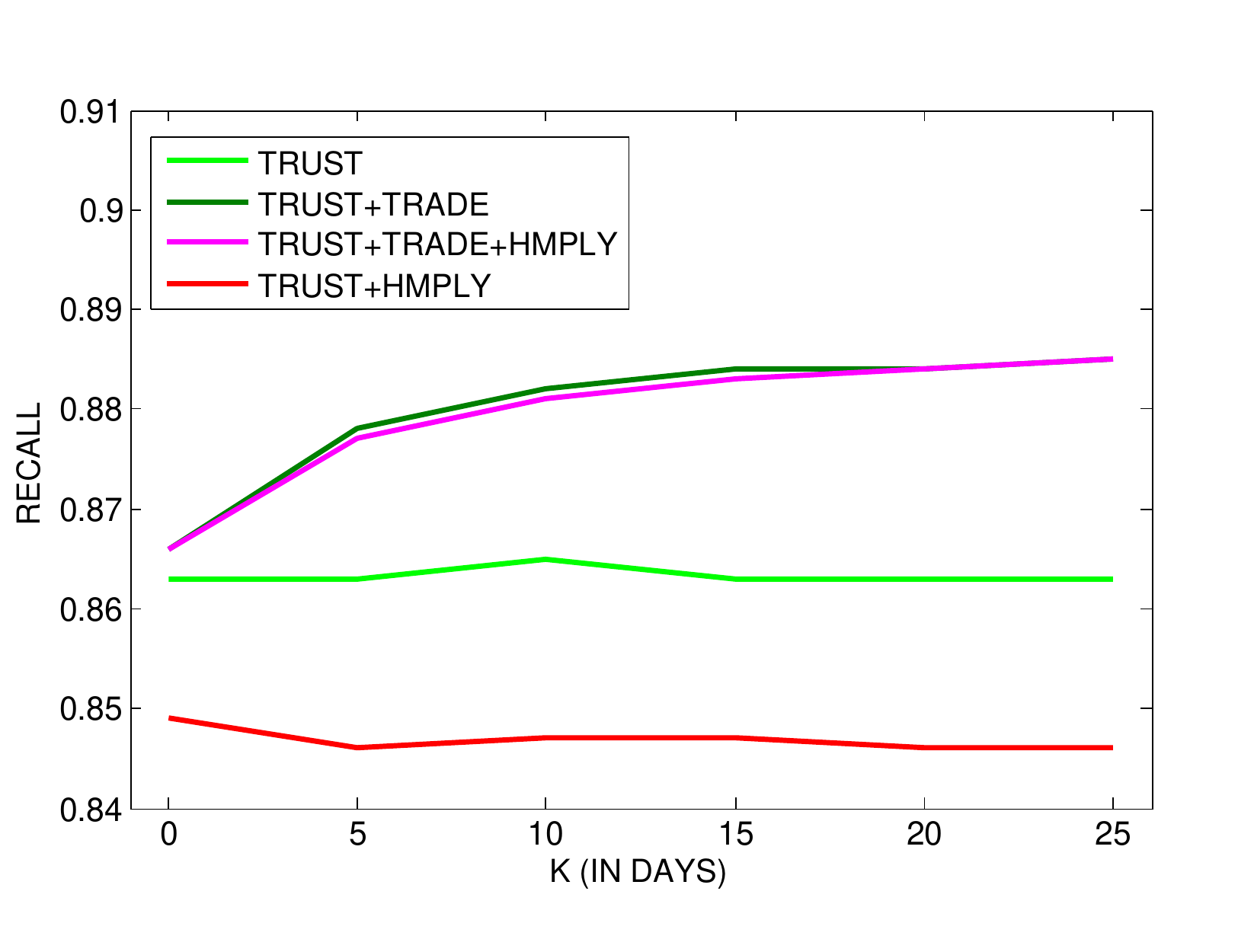}
             }
             
             \caption{Comparing  class weighted accuracy, $F_{1}$ measure, precision, and  recall for trust reciprocation prediction. Here K is the size of time window from the formation of intial trust link between the nodes. The prediction accuracy increases by incorporating trade interactions after the intial trust link was formed. This figure shows that trade features are more significant once a initial trust link is established between two nodes.}\label{fig:measures}
              \vspace*{-2em}
\end{figure*}

 Table~\ref{tab:TableComparingReciprocationAccuracyUsingDifferentFeatureSets}  compares the trust reciprocation accuracy for various  feature sets. The addition of trade features boosts the performance of the predictive model over the case when no trade features are used. All the feature values for this experiment are computed using the values before the trust request was initialted between $A$ and $B$. Thus we do not take into account the impact of future time window (K) in this experiment. The class weighted accuracy (CWA) is highest in case when all the features are used. However, the difference between the CWA using trust and homophily and all features is not very significant. This trend can be observed over all the accuracy measures. However column 2 of the table shows that the AUC measures are higher when using only trust features and AUC drops slightly as we include the trade features. 
 
 Since this table by itself is  not sufficient to draw concrete conclusions, we extend this experiment to include the variation of future time window  size (K) for all features and monitor its impact in term of accuracy of the prediction model.
Figure~\ref{fig:measures} shows the results of this experiment. As mentioned earlier, we study the impact of varying the time windows size (from 0 to 25 days) for all the features. In the previous experiment, we considered the features values without any time window (K=0). As shown in Figure~\ref{fig:measures} and Figure~\ref{fig:AUC}, using trade as an additional feature in the prediction model outperforms the model which uses only trust or trust and homophily only. The performance of the model with trade features  with trust and/or homophily) increase as we increase the size of the time windows from 0 to 25. We also find that addition of homophily features do not have a significant impact in predicting reciprocation in trust network. This is an interesting finding for trust reciprocation prediction because it is a general notion that homophily is significant in predicting trust links \cite{6113234}. A possible explanation of this observation is that the reciprocation phenomenon is significantly different from a normal trust formation phenomenon. As we know, in reciprocation there is already a one sided relationship established and a reciprocation might depend on entirely other dynamics such as trade interactions.

In Figure ~\ref{fig:AUC}, the AUC for the prediction model with trade features actually increase as we increase the time window size (K). Thus this experiment confirms that using trade features boosts the performance of the prediction model for predicting trust reciprocations. One of the reasons why performance of prediction model keeps on increasing as  the time window (K) is increased is due to the fact that players start doing other activities such as trade once a player has initiated a trust request. So the trade activity increases after the trust is initiated by any player and the trust activity pattern helps to discriminate whether a trust reciprocation will happen or not.

In conclusion, the experimental results  confirms our hypothesis that trust reciprocation is significantly improved by incorporating features from other heterogeneous networks such as trade over. We also found that using homophily features does not necessarily improve the accuracy in trust reciprocation prediction. This makes trust reciprocation a all together different phenomenon from trust link formation.